\documentclass{nature}

\usepackage{cite}
\usepackage{amsmath,amssymb,amsfonts}
\usepackage{algorithmic}
\usepackage{graphicx}
\usepackage{textcomp}
\usepackage[dvipsnames]{xcolor}
\title{Nanoscale High Transition Temperature Superconducting Quantum Interference Device Transimpedance Amplifier}

\author{Hao Li$^1$, Ethan Y. Cho$^1$, Han Cai$^1$, Shane A. Cybart$^{1}$}
\makeatletter
\let\saved@includegraphics\includegraphics
\AtBeginDocument{\let\includegraphics\saved@includegraphics}
\renewenvironment*{figure}{\@float{figure}}{\end@float}
\makeatother

\begin{document}

\maketitle
\begin{affiliations}
 \item Department of Electrical and Computer Engineering, University of California, Riverside
\end{affiliations}

\begin{abstract}
As the quantum generation of electronics takes the stage, a cast of important support electronics is needed to connect these novel devices to our classical worlds. In the case of superconducting electronics, this is a challenge because the Josephson junction devices they are based upon require tiny current pulses to create and manipulate the single flux quanta which guide their operation. Difficulty arises in transitioning these signals through large temperature gradients for connection to semiconductor components. In this work, we present nano superconducting quantum interference devices (SQUID) with critical dimensions as small as 10 nm from the high-transition-temperature superconductor YBa$_2$Cu$_3$O$_{7-\delta}$ (YBCO). We integrate these nano-SQUIDs with nano-isolated inductively coupled control lines to create a low power superconducting output driver capable of transimpedance conversion over a very wide temperature range. 
\end{abstract}

\maketitle

Superconductive electronics (SCE) based on Josephson junctions could play a key role in delivering high-speed, low-power, high-performance, centralized and specialized computing. With demonstrated switching speeds of several hundred GHz\cite{chen1999rapid} and energy dissipation on the order of $10^{-19}$ Joules per transition\cite{mukhanov2011energy}, single flux quantum (SFQ) logic\cite{likharev1991rsfq} is one of the most mature and viable post-CMOS technologies. Unlike CMOS, which is based on electric charge, Josephson junctions are quantum devices with operation based on the quantum mechanical wave functions of the superconducting electrons\cite{josephson1962possible}. As a result, SCE can also offer new functionalities beyond conventional computing. For example, quantum computing with superconductors has flourished over the past two decades and could provide big breakthroughs in the factoring of large numbers\cite{devoret2013superconducting}. Quantum annealing processors based on SCE can efficiently solve certain optimization problems\cite{king2018observation}. Neuromorphic computing\cite{shainline2017superconducting} and ultra-low power adiabatic reversible computing\cite{takeuchi2013adiabatic,takeuchi2015thermodynamic} are other newly emerging computational technologies based on SCE.

In all of these SCE technologies, CMOS plays a key role in connecting the inputs and outputs of these cryogenic technologies to our lives at room temperature. SCE and Josephson junctions are low impedance (1-2~$\Omega$) flux-based devices that rely on mA level current pulses for operation. Where CMOS, on the other hand, are voltage devices requiring higher impedance millivolt level signals. Therefore there is a critical need for low power transimpedance amplifiers for matching SCE to room temperature electronics.

Furthermore, it is also highly desirable to operate over a wide cryogenic temperature range to move the power dissipation from 4 K or 10 mK (in the case of quantum computing) to higher temperature such as the upper 40 K cooling stage in typical refrigeration systems to lower the cooling power required for the system. SCE from ceramic high transition temperature (high-$T_C$) superconductors could fill this role, however these materials are notoriously difficult to process and breakthroughs are needed in device fabrication.

The most commonly studied high-$T_C$ superconductor (HTS) is YBa$_2$Cu$_3$3O$_{7-\delta}$(YBCO). It crystallizes in a distorted perovskite orthorhombic structure. As a result the electrical transport is highly anisotropic and varies in every direction. The resistivty along the \textit{b} direction is the lowest ($\sim$~100~$\mu\Omega\cdot$cm) and is 3 and 10 times higher along the \textit{a} and \textit{c} directions respectively\cite{friedmann1990direct}. Similarly the superconducting coherence length, $\xi$, is very short and anisotropic. Along the \textit{a} crystalline direction $\xi_a$ is only 2 nm but even shorter  along the \textit{b} and \textit{c} directions\cite{worthington1987anisotropic}.
The poor transport properties along the \textit{c} direction coupled with the fact that the highest-quality films have their c-axes' orientated perpendicular to the substrate motivates Josephson junctions with geometries where current flows in the \textit{a}-\textit{b} plane. Furthermore, the fabrication of vias and crossovers for the creation of multi-layer input coils and complex layouts are also very challenging and typically involve patterning and overlapping films etched at angles to make connections in the \textit{a}-\textit{b} plane\cite{kingston1990multilayer,van1996ceo2}. YBCO is very sensitive to heat and process damage which makes physical etching of feature sizes less than a few micrometers very challenging\cite{kimura2009operation}. 
The recent development of focused helium ion beam (FHIB) direct-write nano-lithography of HTS materials opens up the door to novel devices with nanoscale geometries\cite{cybart2015nano} because it does not involve removal or etching of material. 
In this work, we overcome many of the fabrication challenges associated with HTS to develop a high-transition temperature nano superconducting quantum interference device (SQUID) flux-to-voltage transimpedance amplifier that can operate over a broad range of parameters and temperatures. Our device lies entirely within the \textit{a-b} plane of a single YBCO film, and occupies an area less than 1 $\mu$m$^2$. The heart of our device is a high impedance (30~$\Omega$) nano-SQUID. Nano-SQUIDs are robust to stray magnetic fields due to their very small areas, however this property also makes the coupling of flux into the SQUID from on-chip lines very difficult.  In practice, it is hard to inductively couple magnetic flux into a nano-SQUID because the area is so small, a global large external magnetic field hundreds of times stronger than Earth's field is usually needed. In our design, we use FHIB direct-write lithography to place a control line just a few nanometers away from the SQUID. This close proximity allows for strong coupling because the field strength $B$ is inversely proportional to the distance $r$ from the current carrying wire, $B\sim1/r$. 

Nano-patterning of YBCO is accomplished using FHIB direct-write lithography. This technology utilizes a finely focused 0.5~nm diameter helium ion beam from a gas field ion source\cite{ward2006helium}. The irradiation causes disorder in the YBCO lattice that at moderate doses (10$^{15}$~ions/cm$^2$) converts the material from superconductor to insulator\cite{white1988controllable,lang2006ion}. With this method, very fine sub-10~nm insulating lines can be written directly into the plane of a YBCO thin film to create Josephson junctions\cite{cybart2015nano,muller2019josephson,cho2018direct,cho2018superconducting}. To fabricate nano-SQUID transimpedance amplifiers, we began with 5~mm $\times$ 5~mm sapphire substrates commercially coated with a 35-nm-thick YBCO thin-film grown by reactive coevaporation\cite{kinder1997ybco}. The film was capped \textit{in-situ} with a thermally evaporated 200-nm-thick Au layer for electrical contacts. Laser photolithography and Ar ion milling were used to pattern large circuit features into both the Au and YBCO layers. A second photolithography and Au chemical etch were performed to uncover the YBCO film in the centre of the substrate for FHIB direct-writing as shown in Fig.~1a. The centre region was designed to contain four test structures, each containing a nano-SQUID integrated with a strongly coupled on-chip magnetic flux control line. Fig.~1b is a photograph of a representative device which shows a zoomed in view of the YBCO electrodes for a nano-SQUID, an on-chip control current path and illustrations where direct-write patterning was performed. Irradiation with a dose of $\sim$8$\times$10$^{17}$~He$^+$/cm$^2$ was used to modify the film to form the SQUID loop (white square) and the insulating lines (white lines) which define the on-chip control line. Using a smaller ion dose of 6$\sim$8~$\times$10$^{16}$ He$^+$/cm$^2$, a very narrow insulating barrier (of the order of $\xi$) for YBCO Josephson junctions (red lines) was also created.

The FHIB nano-SQUIDs are not visible using optical microscopy because of their small size and the fact that no materials have been removed in the direct writing process. However, excellent contrast between insulating and conducting regions can be observed with helium ion microscopy. Fig.~1c shows a scanning helium ion image of a nano-SQUID device after implantation. The altered electrical properties provide sufficient contrast to discern the nano-SQUIDs fabricated by FHIB. We also remark that regions where high dosages of helium have been delivered, can also be observed by atomic force microscopy (AFM), as shown in Fig.~1d. This is consistent with the findings of previous work involving helium irradiated silicon\cite{livengood2009subsurface}. The high energy helium ions (or helium atoms after recombining with electrons and becoming neutral) become trapped in the substrate. When exceeding a dose threshold the helium coalesces and causes a local expansion or bubble that raises the exposed area. In the case of Fig.~1d, application of a helium irradiation dose of $\sim$8~$\times$10$^{17}$~He$^+$/cm$^2$ to a YBCO on sapphire substrate changes the average height by $\sim$12 nm, or $\sim$30$\%$ of the thickness of the 35 nm thick YBCO film. The mechanical distortion together with the structural defects such as oxygen displacement and amorphization of the crystal likely contribute to the insulating properties of the film after high dose irradiation.

Several FHIB nano-SQUIDs with integrated control lines were written with square SQUID loops with side ranging from 900 nm to 10 nm  with (300 $\times$ 2 nm) Josephson junctions to determine the optimum size for our transimpedance amplifier. After fabrication, \textit{I-V} characteristics were measured in a liquid helium, dip probe with a three-layer magnetic shield. Subsequently, devices were DC biased slightly over critical current $I_\text{C}$($\sim$1.2$I_\text{C}$), and a current $I_\text{M}$ was swept through the on-chip control line, which generated a relative strong magnetic field, and the voltage across the device was measured to examine $V$-$I_\text{M}$. The process was found to be reliable and reproducible. The data recorded for all of these devices is included in supplemental Fig. S1. 

Helium ion images of the two smallest nano-SQUIDs with 10~nm and 50~nm square loops are shown in Fig. 2. These SQUIDs have the smallest loops demonstrated to date in any superconducting material. The \textit{I-V} curves for these smallest nano-SQUIDs are shown in Fig.~3a. They exhibit relatively high resistance of 25 $\Omega$ and 32 $\Omega$ for the 50 nm and 10 nm nano-SQUIDs respectively. Because the $I_\text{C}R_\text{N}$ product for a material is approximately constant, higher resistance implies lower $I_\text{C}$. Smaller $I_\text{C}$ is desirable for a given geometry because it leads to smaller values of the inductance parameter $\beta_L=I_\text{C}L/\Phi_0$, where \textit{L} is the inductance of nano-SQUID loop. This in-turn results in larger amplitude of the SQUIDs voltage modulation as shown in Fig.~3b. The amplitude of the voltage for these nano-SQUIDs reached maximums of 0.8 and 0.5 mV for the 50 and 10 nm nano-SQUIDs respectively. This is many times larger than typical values of nano-SQUID devices found in the literature.

While these smallest devices are an advance in scaling down the size of nano-SQUIDs, unfortunately, the inductive coupling strength of our integrated control line was too weak for operation of the nano-SQUID as a transimpedance amplifier. Our further studies in this effort focused on devices with 400-nm square SQUID loops because they exhibited the best combination of output voltage and inductive coupling between the control line. Current-voltage  (\textit{I-V}) characteristics for a 400-nm square loop device are shown in Fig.~4a from 4.2 K up to 52 K. The critical current \textit{I}$_C$ ranged from 50~$\mu$A to 5~$\mu$A, while a nearly temperature independent voltage-state resistance \textit{R}$_N$ was observed to be 30~$\Omega$. This desirable feature allows us to operate the SQUID over a wide temperature range with a stable output impedance. The voltage magnetic field characteristics (Fig.~4b) for this SQUID are nearly ideal with excellent uniformity, and symmetry. They exhibit a periodicity in external magnetic field with a period $\Delta B\sim$~3.6 mT. The amplitude of the voltage modulation ranges from 0.5 mV to 0.03 mV with temperature from 4.2 K up to 52 K. The large amplitude and wide operating temperature range show the capability of these devices to output large voltages at high and low temperatures.  

The local minimal points in the voltage-magnetic field (\textit{V-B}) curves in Fig.~3b represent integer numbers of the single flux quantum, $\Phi_0$, contained within the SQUID. An area can be defined as $A_\text{eff} \equiv \Phi_0/\Delta B$. In practice this typically doesn't match the geometric area for a SQUID due to the focusing of flux into the SQUID from nearby electrodes. With the inspiration from the previous work about planar junctions\cite{rosenthal1991flux}, we estimated the effective area to include the points whose nearest non-superconducting edges are the four sides of the insulating white square, as shown in Fig.~1a. This estimate provides an area of 0.56~$\mu$m$^2$ that is in good agreement with $A_\text{eff}$.

We measured the noise properties of the nano-SQUID by using a open loop method. The voltage noise was recorded by spectrum analyzer, when a bias current $\sim$60~$\mu$A and flux bias -0.13 $\Phi_0$ were used to achieve the steepest slope in the \textit{V-B} curve. Dividing the voltage noise by the corresponding transfer function yielded a flux noise of $\sim$~300~n$\Phi_0$/Hz$^{1/2}$ at 100 kHz at 4.2 K, as shown in supplemental Fig. S2. This value is typical of nano-SQUIDs\cite{arpaia2016improved,chen2016high} measured by this method, but larger than that of the state of art high-$T_c$ and low-$T_c$ nano-SQUIDs\cite{schwarz2015low, vasyukov2013scanning} measured with the help of low-temperature amplifiers and or flux-locked loops employing bias reversal for elimination of critical current noise.

   

To characterize the coupling of the nano-isolated control line to the 400-nm square loop SQUID a current was swept through the control line and the voltage response of the SQUID was measured (Fig.~5a), for several different values of SQUID DC bias currents ($\sim$8~$\mu$A to $\sim$17.5~$\mu$A). The shape of the characteristics and maximum amplitude of the modulation voltage $V_\text{max}$ were found to be in good agreement with that measured using externally applied magnetic field (Fig. 4b). The mutual inductance \textit{M}, between the on-chip control line and the main body of the nano-SQUID was calculated to be 0.067~pH by the relation $M = \Phi_0/\Delta I$, where $\Delta I$ is the amount of current needed to couple a single flux quantum into the nano-SQUID. The period of the control current, $\sim$31~mA (Fig.~5a) corresponds to approximately a field strength of 3.6~mT (Fig. 4b) or a transfer factor of 8.6 mA/mT.   
For operation as a transimpedance output driver the SQUID was DC biased at $\sim$13$\mu$A and a flux of 0.1$\Phi_0$ was applied to field bias the SQUID to the maximum transfer factor. A 3~mA input current was then pulsed through the on-chip control line generating an output voltage pulse of $\sim$~0.35 mV, resulting in a transimpedance $\sim$0.1~$\Omega$ as shown in Fig.~5b. Because the control line is completely superconducting no power is dissipated in its operation. Power is dissipated in the junctions of the SQUID when they are in the voltage state but it is estimated to be only 9 nW due to the small bias current of the SQUID $\sim$13$\mu$A, compared to the power dissipation 3.1 mW of the recent developed cryogenic semiconductor trans-impedance amplifier\cite{bonen2018cryogenic}.

Our findings demonstrate the nano-SQUID as a promising low power trans-impedance amplifier for connection of superconducting circuits to room temperature silicon electronics.  Gas field ion source direct-write patterning of HTS has given us the ability to aggressively scale and integrate device features in superconducting electronics. The integration of nano-scale Josephson junctions, SQUID loops and in-plane insulators for control lines to build and test a trans-impedance amplifier is an example of the power of this technique. The possibilities for other SCE circuit implementations are countless as we move into a new era in HTS electronics.



\begin{methods}

\subsection{Sample fabrication}

A 35-nm-thick YBCO thin-film was grown on CeO$_2$ buffered r-plane sapphire by reactive coevaporation by Ceraco GmbH. After deposition a 200 nm thick gold film is thermally evaporated \textit{in-situ}. The wafers were coated with a thick layer of photo resist and diced into 5 mm $\times$ 5 mm chips along the YBCO a-b axis with a Disco DAD321 automatic dicing saw. The substrates were spin coated with Fuji film OCG 825 resist at 5000 rpm, and baked for 60~s on a 90$^\circ$C hotplate. A Microtech LW405 maskless laser direct-write photolithography system was used to expose the resist with a 405 nm wavelength laser in raster scanning mode. After exposure, development was performed using Fuji film OCG 934 for 90 s. Subsequently, the pattern was etched into both the Au and YBCO layers in a 21 cm Ion Tech argon ion mill with power 150 W, Ar pressure 1.7E-4 torr and energy 500 eV.  A second photolithography step was performed to mask the electrical contacts and Transene gold etchant was used to expose the YBCO film for FHIB direct writing. The sample was loaded into a Carl Zeiss Orion Plus Helium Ion Microscope. A 0.5-nm-diameter, 32-keV focused helium ion beam was formed by aligning the atomically sharp tungsten tip with the help of the electrical extracting, accelerating, and focusing system. The custom-defined vector pattern (direct-writing) was generated by NanoPatterning and Visualization Engine (NPVE). For Josephson junctions the single line irradiation dose ~600 ions/nm was used with a dwell time of 1~$\mu$s, a point spacing of 0.25 nm. For the SQUID loop, and other insulators an area irradiation dose was used ~3200 ions/nm$^2$ with spacing 0.25 nm in both x-axis and y-axis. 

\subsection{Electrical transport measurement}

The chip was mounted with Devcon epoxy 14270 and wedge wire bonded to a 44-pin copper plate chip carrier with a Lake Shore silicon diode DT-670 to ensure the accurate monitoring of the local temperature of the chip. The whole chip carrier was mounted on a vacuum tight dip probe with twisted pair wires and pin-type low pass EMI filter with bandwidth 1 MHz. The test signal was generated by an Agilent arbitrary waveform generator A33522, which was conditioned through a custom made iso-amplifier.  The output voltage of the nano-SQUID was amplified by a Stanford Research preamplifier SR560. The data was recorded by a National Instruments analog ot digital converter NI9215 connected to a PC. For noise measurements the chip was mounted on a G10 fiberglass dip probe to avoid any noise from normal metals. A niobium can and cryoperm shield was used to reduce the background magnetic noise. Lead acid batteries were used for the bias current and control current for the on-chip coils. The bias current was also filtered by a second order low pass RC filter working at low temperature. The output voltage was amplified by a SR560 with a white noise ~4 nV/Hz-1/2, then recorded by a Hewlett Packard spectrum analyzer 3562A.
  
\end{methods}


\bibliography{sample}

\begin{thebibliography}{10}
\expandafter\ifx\csname url\endcsname\relax
  \def\url#1{\texttt{#1}}\fi
\expandafter\ifx\csname urlprefix\endcsname\relax\def\urlprefix{URL }\fi
\providecommand{\bibinfo}[2]{#2}
\providecommand{\eprint}[2][]{\url{#2}}

\bibitem{chen1999rapid}
\bibinfo{author}{Chen, W.}, \bibinfo{author}{Rylyakov, A.},
  \bibinfo{author}{Patel, V.}, \bibinfo{author}{Lukens, J.} \&
  \bibinfo{author}{Likharev, K.}
\newblock \bibinfo{title}{Rapid single flux quantum t-flip flop operating up to
  770 ghz}.
\newblock \emph{\bibinfo{journal}{IEEE Transactions on Applied
  Superconductivity}} \textbf{\bibinfo{volume}{9}}, \bibinfo{pages}{3212--3215}
  (\bibinfo{year}{1999}).

\bibitem{mukhanov2011energy}
\bibinfo{author}{Mukhanov, O.~A.}
\newblock \bibinfo{title}{Energy-efficient single flux quantum technology}.
\newblock \emph{\bibinfo{journal}{IEEE Transactions on Applied
  Superconductivity}} \textbf{\bibinfo{volume}{21}}, \bibinfo{pages}{760--769}
  (\bibinfo{year}{2011}).

\bibitem{likharev1991rsfq}
\bibinfo{author}{Likharev, K.~K.} \& \bibinfo{author}{Semenov, V.~K.}
\newblock \bibinfo{title}{Rsfq logic/memory family: A new josephson-junction
  technology for sub-terahertz-clock-frequency digital systems}.
\newblock \emph{\bibinfo{journal}{IEEE Transactions on Applied
  Superconductivity}} \textbf{\bibinfo{volume}{1}}, \bibinfo{pages}{3--28}
  (\bibinfo{year}{1991}).

\bibitem{josephson1962possible}
\bibinfo{author}{Josephson, B.~D.}
\newblock \bibinfo{title}{Possible new effects in superconductive tunnelling}.
\newblock \emph{\bibinfo{journal}{Physics letters}}
  \textbf{\bibinfo{volume}{1}}, \bibinfo{pages}{251--253}
  (\bibinfo{year}{1962}).

\bibitem{devoret2013superconducting}
\bibinfo{author}{Devoret, M.~H.} \& \bibinfo{author}{Schoelkopf, R.~J.}
\newblock \bibinfo{title}{Superconducting circuits for quantum information: an
  outlook}.
\newblock \emph{\bibinfo{journal}{Science}} \textbf{\bibinfo{volume}{339}},
  \bibinfo{pages}{1169--1174} (\bibinfo{year}{2013}).

\bibitem{king2018observation}
\bibinfo{author}{King, A.~D.} \emph{et~al.}
\newblock \bibinfo{title}{Observation of topological phenomena in a
  programmable lattice of 1,800 qubits}.
\newblock \emph{\bibinfo{journal}{Nature}} \textbf{\bibinfo{volume}{560}},
  \bibinfo{pages}{456} (\bibinfo{year}{2018}).

\bibitem{shainline2017superconducting}
\bibinfo{author}{Shainline, J.~M.}, \bibinfo{author}{Buckley, S.~M.},
  \bibinfo{author}{Mirin, R.~P.} \& \bibinfo{author}{Nam, S.~W.}
\newblock \bibinfo{title}{Superconducting optoelectronic circuits for
  neuromorphic computing}.
\newblock \emph{\bibinfo{journal}{Physical Review Applied}}
  \textbf{\bibinfo{volume}{7}}, \bibinfo{pages}{034013} (\bibinfo{year}{2017}).

\bibitem{takeuchi2013adiabatic}
\bibinfo{author}{Takeuchi, N.}, \bibinfo{author}{Ozawa, D.},
  \bibinfo{author}{Yamanashi, Y.} \& \bibinfo{author}{Yoshikawa, N.}
\newblock \bibinfo{title}{An adiabatic quantum flux parametron as an
  ultra-low-power logic device}.
\newblock \emph{\bibinfo{journal}{Superconductor Science and Technology}}
  \textbf{\bibinfo{volume}{26}}, \bibinfo{pages}{035010}
  (\bibinfo{year}{2013}).

\bibitem{takeuchi2015thermodynamic}
\bibinfo{author}{Takeuchi, N.}, \bibinfo{author}{Yamanashi, Y.} \&
  \bibinfo{author}{Yoshikawa, N.}
\newblock \bibinfo{title}{Thermodynamic study of energy dissipation in
  adiabatic superconductor logic}.
\newblock \emph{\bibinfo{journal}{Physical Review Applied}}
  \textbf{\bibinfo{volume}{4}}, \bibinfo{pages}{034007} (\bibinfo{year}{2015}).

\bibitem{friedmann1990direct}
\bibinfo{author}{Friedmann, T.}, \bibinfo{author}{Rabin, M.},
  \bibinfo{author}{Giapintzakis, J.}, \bibinfo{author}{Rice, J.} \&
  \bibinfo{author}{Ginsberg, D.}
\newblock \bibinfo{title}{Direct measurement of the anisotropy of the
  resistivity in the a-b plane of twin-free, single-crystal, superconducting
  yba 2 cu 3 o 7- $\delta$}.
\newblock \emph{\bibinfo{journal}{Physical Review B}}
  \textbf{\bibinfo{volume}{42}}, \bibinfo{pages}{6217} (\bibinfo{year}{1990}).

\bibitem{worthington1987anisotropic}
\bibinfo{author}{Worthington, T.~K.}, \bibinfo{author}{Gallagher, W.} \&
  \bibinfo{author}{Dinger, T.}
\newblock \bibinfo{title}{Anisotropic nature of high-temperature
  superconductivity in single-crystal y 1 ba 2 cu 3 o 7- x}.
\newblock \emph{\bibinfo{journal}{Physical review letters}}
  \textbf{\bibinfo{volume}{59}}, \bibinfo{pages}{1160} (\bibinfo{year}{1987}).

\bibitem{kingston1990multilayer}
\bibinfo{author}{Kingston, J.~J.}, \bibinfo{author}{Wellstood, F.~C.},
  \bibinfo{author}{Lerch, P.}, \bibinfo{author}{Miklich, A.~H.} \&
  \bibinfo{author}{Clarke, J.}
\newblock \bibinfo{title}{Multilayer yba2cu3o x-srtio3-yba2cu3o x films for
  insulating crossovers}.
\newblock \emph{\bibinfo{journal}{Applied Physics Letters}}
  \textbf{\bibinfo{volume}{56}}, \bibinfo{pages}{189--191}
  (\bibinfo{year}{1990}).

\bibitem{van1996ceo2}
\bibinfo{author}{Van~Wijck, M.} \emph{et~al.}
\newblock \bibinfo{title}{Ceo2 as insulation layer in high t c superconducting
  multilayer and crossover structures}.
\newblock \emph{\bibinfo{journal}{Applied physics letters}}
  \textbf{\bibinfo{volume}{68}}, \bibinfo{pages}{553--555}
  (\bibinfo{year}{1996}).

\bibitem{kimura2009operation}
\bibinfo{author}{Kimura, T.} \emph{et~al.}
\newblock \bibinfo{title}{Operation of toggle flip-flop circuits up to 500 ghz
  based on vertically-stacked high-temperature superconductor josephson
  junctions}.
\newblock \emph{\bibinfo{journal}{IEEE Transactions on Applied
  Superconductivity}} \textbf{\bibinfo{volume}{19}}, \bibinfo{pages}{127--130}
  (\bibinfo{year}{2009}).

\bibitem{cybart2015nano}
\bibinfo{author}{Cybart, S.~A.} \emph{et~al.}
\newblock \bibinfo{title}{Nano josephson superconducting tunnel junctions in
  {YBa$_2$Cu$_3$O$_{7-\delta}$} directly patterned with a focused helium ion
  beam}.
\newblock \emph{\bibinfo{journal}{Nat. Nanotechnol.}}
  \textbf{\bibinfo{volume}{10}}, \bibinfo{pages}{598} (\bibinfo{year}{2015}).

\bibitem{ward2006helium}
\bibinfo{author}{Ward, B.}, \bibinfo{author}{Notte, J.~A.} \&
  \bibinfo{author}{Economou, N.}
\newblock \bibinfo{title}{Helium ion microscope: A new tool for nanoscale
  microscopy and metrology}.
\newblock \emph{\bibinfo{journal}{Journal of Vacuum Science \& Technology B:
  Microelectronics and Nanometer Structures Processing, Measurement, and
  Phenomena}} \textbf{\bibinfo{volume}{24}}, \bibinfo{pages}{2871--2874}
  (\bibinfo{year}{2006}).

\bibitem{white1988controllable}
\bibinfo{author}{White, A.~E.} \emph{et~al.}
\newblock \bibinfo{title}{Controllable reduction of critical currents in
  yba2cu3o7- $\delta$ films}.
\newblock \emph{\bibinfo{journal}{Applied physics letters}}
  \textbf{\bibinfo{volume}{53}}, \bibinfo{pages}{1010--1012}
  (\bibinfo{year}{1988}).

\bibitem{lang2006ion}
\bibinfo{author}{Lang, W.} \emph{et~al.}
\newblock \bibinfo{title}{Ion-beam direct-structuring of high-temperature
  superconductors}.
\newblock \emph{\bibinfo{journal}{Microelectron. Eng.}}
  \textbf{\bibinfo{volume}{83}}, \bibinfo{pages}{1495--1498}
  (\bibinfo{year}{2006}).

\bibitem{muller2019josephson}
\bibinfo{author}{M\"uller, B.} \emph{et~al.}
\newblock \bibinfo{title}{Josephson junctions and {SQUIDs} created by focused
  helium-ion-beam irradiation of
  {${\mathrm{Y}\mathrm{Ba}}_{2}{\mathrm{Cu}}_{3}{\mathrm{O}}_{7}$}}.
\newblock \emph{\bibinfo{journal}{Phys. Rev. Appl.}}
  \textbf{\bibinfo{volume}{11}}, \bibinfo{pages}{044082}
  (\bibinfo{year}{2019}).
\newblock
  \urlprefix\url{https://link.aps.org/doi/10.1103/PhysRevApplied.11.044082}.

\bibitem{cho2018direct}
\bibinfo{author}{Cho, E.~Y.} \emph{et~al.}
\newblock \bibinfo{title}{Direct-coupled micro-magnetometer with {Y-Ba-Cu-O}
  nano-slit squid fabricated with a focused helium ion beam}.
\newblock \emph{\bibinfo{journal}{Appl. Phys. Lett.}}
  \textbf{\bibinfo{volume}{113}} (\bibinfo{year}{2018}).
\newblock \urlprefix\url{http://par.nsf.gov/biblio/10076662}.

\bibitem{cho2018superconducting}
\bibinfo{author}{Cho, E.~Y.}, \bibinfo{author}{Zhou, Y.~W.},
  \bibinfo{author}{Cho, J.~Y.} \& \bibinfo{author}{Cybart, S.~A.}
\newblock \bibinfo{title}{Superconducting nano josephson junctions patterned
  with a focused helium ion beam}.
\newblock \emph{\bibinfo{journal}{Appl. Phys. Lett.}}
  \textbf{\bibinfo{volume}{113}}, \bibinfo{pages}{022604}
  (\bibinfo{year}{2018}).

\bibitem{kinder1997ybco}
\bibinfo{author}{Kinder, H.} \emph{et~al.}
\newblock \bibinfo{title}{{YBCO} film deposition on very large areas up to
  20$\times$ 20 cm$^2$}.
\newblock \emph{\bibinfo{journal}{Physica C}} \textbf{\bibinfo{volume}{282}},
  \bibinfo{pages}{107--110} (\bibinfo{year}{1997}).

\bibitem{livengood2009subsurface}
\bibinfo{author}{Livengood, R.}, \bibinfo{author}{Tan, S.},
  \bibinfo{author}{Greenzweig, Y.}, \bibinfo{author}{Notte, J.} \&
  \bibinfo{author}{McVey, S.}
\newblock \bibinfo{title}{Subsurface damage from helium ions as a function of
  dose, beam energy, and dose rate}.
\newblock \emph{\bibinfo{journal}{Journal of Vacuum Science \& Technology B:
  Microelectronics and Nanometer Structures Processing, Measurement, and
  Phenomena}} \textbf{\bibinfo{volume}{27}}, \bibinfo{pages}{3244--3249}
  (\bibinfo{year}{2009}).

\end{thebibliography}


\begin{addendum}
 \item[Authors contribution] H.L. and S.C. conceived and planned the experiments. H.C designed and prepared the chips by using laser photolithography, argon ion milling and gold etching.  H.L, and E.Y.C. made the chips by using helium ion microscope. H.L., H.C., and E.Y.C carried out the measurement and characterization at low temperature. H.L, E.Y.C. and S.C. contributed to the interpretation of the results. S.C. took the lead in writing the manuscript.
 
 \item The authors thank Yan-Ting Wang and Stephen McCoy for the morphological analysis of the sample by using atomic force morphology and scanning electronic morphology, Shozo Yoshizumi of Quantum Design for assistance with PPMS measurements, and Miranda Vinay for help with proofreading the manuscript. This work was supported by the Air Force Office of Scientific Research under Grants FA955015-1-0218, and FA9550-19-C-0001; the National Science Foundation under Grant 1664446; the University of California Office of the President, Multi-campus Research Programs and Initiatives under Award No.009556-002; and by the Army Research Office Grant W911NF1710504. 
\item[Competing Interests] The authors declare that they have no
competing financial interests.
 \item[Correspondence] Correspondence and requests for materials
should be addressed to Shane Cybart~(email: cybart@ucr.edu).
\end{addendum}

\begin{figure*}
    {\includegraphics[width = 1\linewidth]{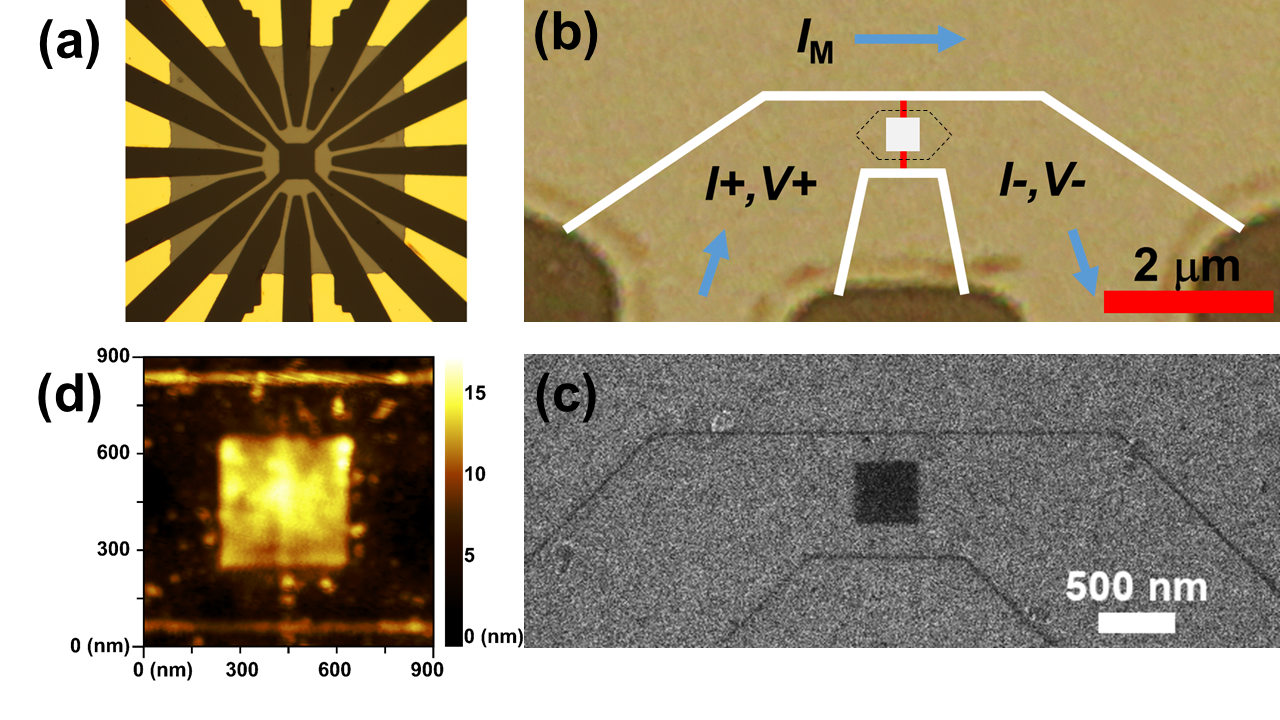}\label{1a}}\hfill
    \caption{(a) Optical image of the central structure of nano-SQUID chips. (b) Zoomed in view of the optical image of the chips depicted where a nano-SQUID was made, the red line represents the barrier of the junction and the white lines and square represented the insulating YBCO, the dashed lines indicate the effective magnetic capture area. (c) Helium ion microscopic image of a YBCO nano-SQUID with a 400-nm wide square loop. The width of junctions is 200 nm.(d) Atomic Force Microscope image of the YBCO nano-SQUID. Helium ions penetrate the YBCO thin film and collect in the substrate. Accumulation of helium causes localized expansion in the substrate that mechanically raises the film.}
    \label{fig:fig1}
\end{figure*}

\begin{figure}
    \centering
    \includegraphics[width=3 in]{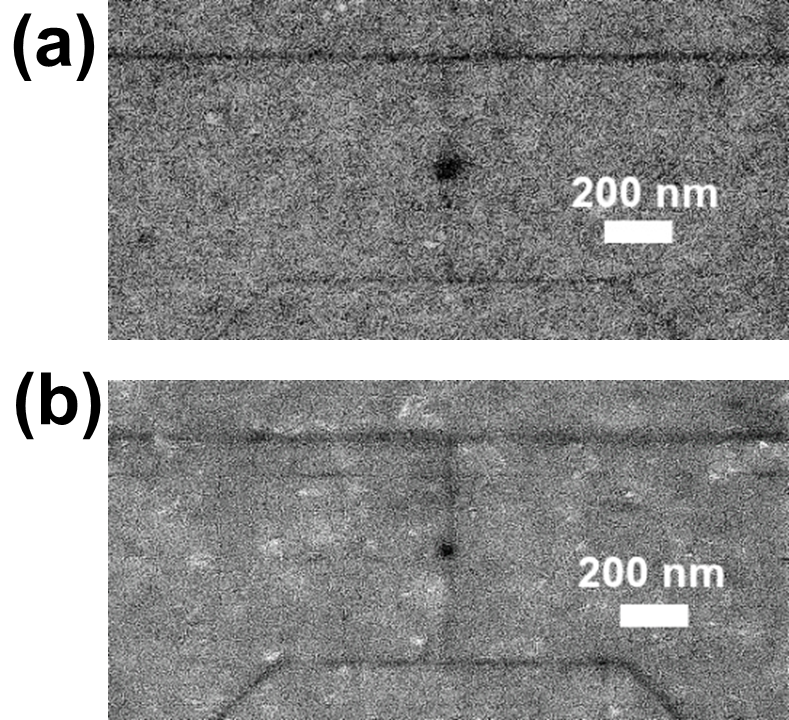}
     \caption{Scanning helium ion image of nano-SQUIDs with 50 nm (a) and 10 nm (b) square SQUID loops. The darker areas represent insulating regions embeded into the plane of a single film. No material is milled or etched. }
    \label{fig:fig2a}
\end{figure}

\begin{figure}
    \centering
    \includegraphics[width=3 in]{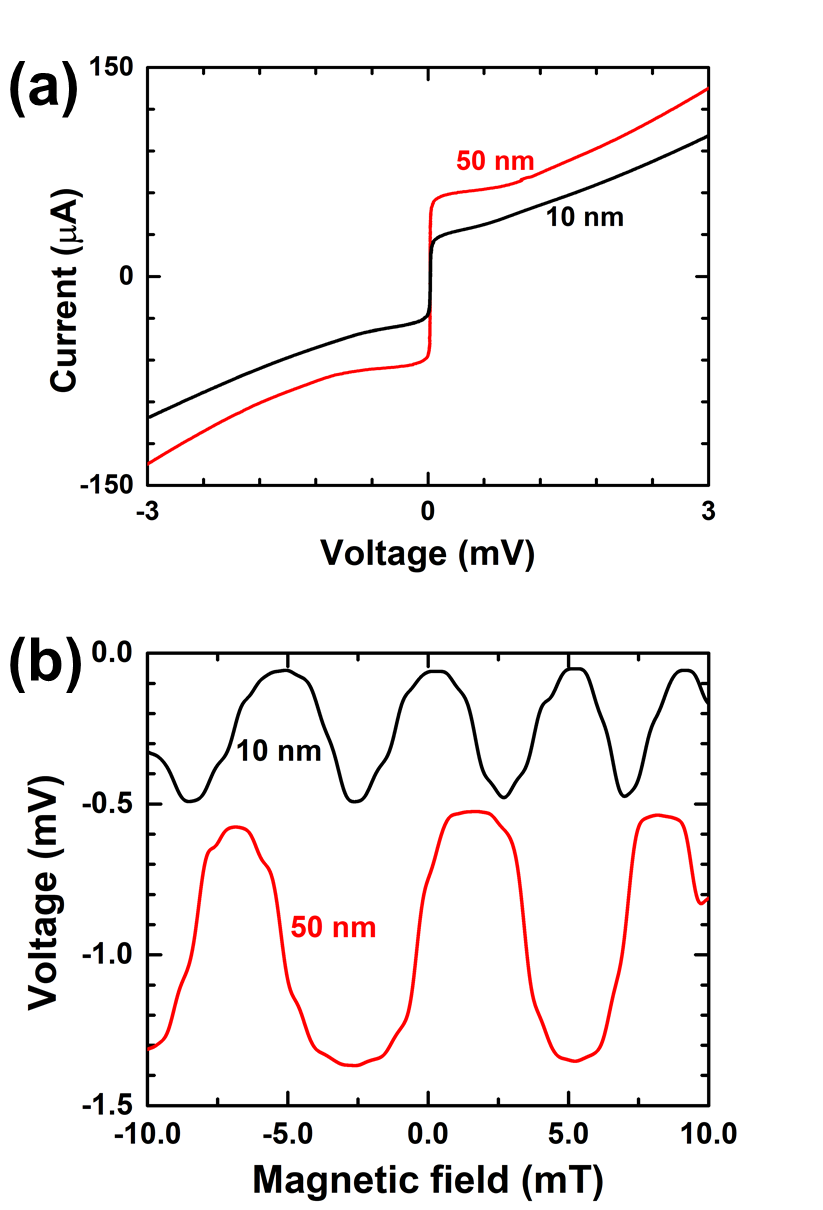}
          \caption{Electrical transport properties of 50-nm and 10-nm wide square loop nano-SQUIDs.  \textit{I-V} characteristics (a) and (b) \textit{V-B} curves of measured at 4.2 K.}
    \label{fig:fig2a}
\end{figure}

\begin{figure}
    \centering
    \includegraphics[width=3 in]{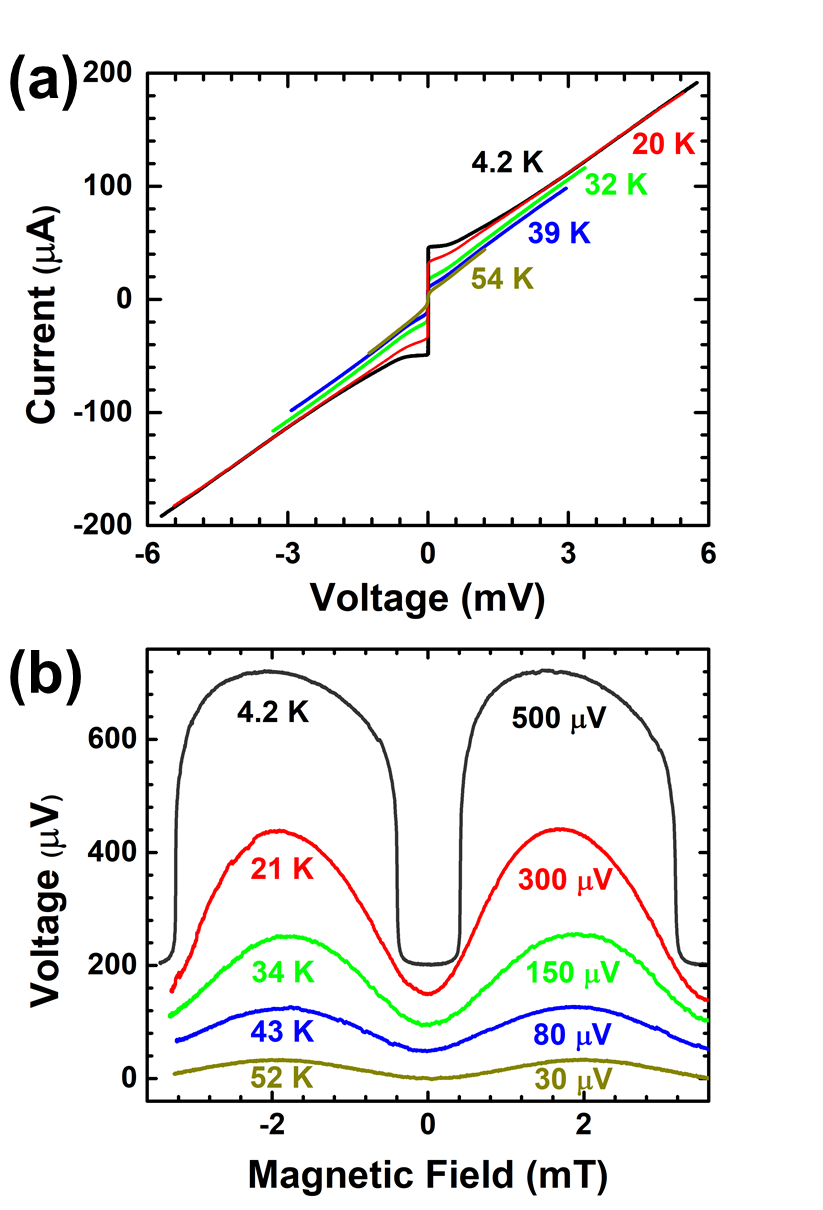}
       \caption{(a) Current-Voltage (\textit{I-V}) characteristics of the nano-SQUIDs consisting of two 200-nm wide Josephson junctions and a 400-nm wide square SQUID loop at five different temperatures; 4.2 K, 20 K, 32 K, 39 K and 54 K. $I_\text{C}$ decreased from 50~$\mu$A to 5 $\mu$A, while the $R_N$ remained at approximately kept at 30 $\Omega$ (b) Voltage-Magnetic Field (\textit{V-B}) curves of the nano-SQUIDs from 4.2 K to 52 K, the modulation voltage decreased from 0.5 mV to 0.03 mV.}
    \label{fig:fig2a}
\end{figure}

\begin{figure}
    \centering
    \includegraphics[width=3 in]{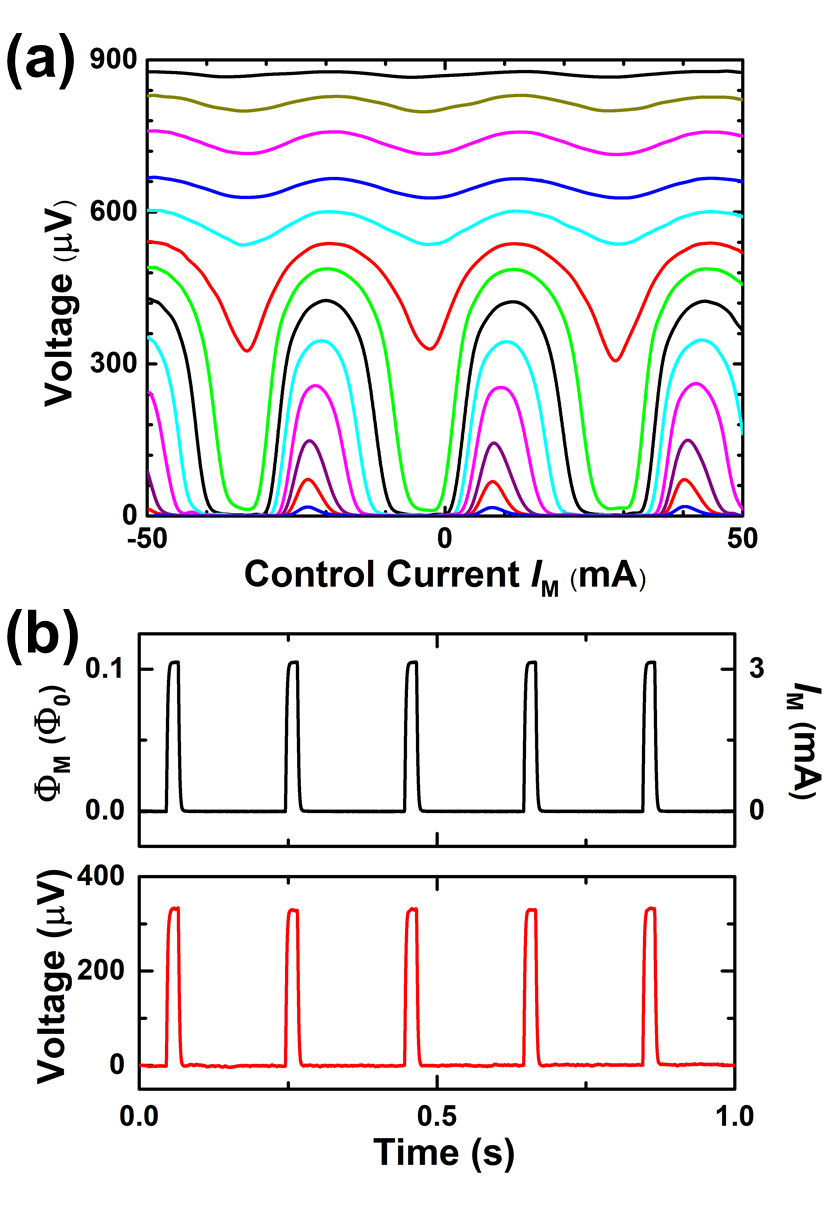}
   \caption{(a) Voltage modulation for the 400-nm square loop nano-SQUID measured with an on-chip nano-isolated control current at different SQUID DC bias currents from $\sim$8~$\mu$A to $\sim$17.5~$\mu$A at 4.2 K. The period of the current $\sim$31 mA represents the generation of a flux equal to $\Phi_0$ in the SQUID. (b) Operation of the nano-SQUID as a transimpedance amplifier. An input current of 3~mA was pulsed through the control line and the corresponding output voltage sequence is shown.}
    \label{fig:fig2a}
   
\end{figure}


\end{document}


\maketitle

\begin{affiliations}
 \item Department of Electrical and Computer Engineering, University of California, Riverside
\end{affiliations}

\subsection{Reproducibility and Scalability}

The nano-SQUIDs with varying size of square SQUID loops ranging from 900 nm to 100 nm were constructed for systematic investigation. The current-voltage (\textit{I-V}) characteristics at 4.2 K are shown in Fig. S1a, showing the critical currents $I_\text{Cs}$ of $\sim$5~$\mu$A and normal resistances $R_\text{Ns}$ ranging from 60 $\Omega$ to 200 $\Omega$. The dependence of the voltage on the control current going through the on-chip coil of the nano-SQUIDs are shown in Fig. S1b. Some of the curvs on y-axis were offset to avoid overlap in the figure. The voltage modulation and $I_\text{C}R_\text{N}$ product of the nano-SQUIDs have a strong positive relation, as presented in Fig.~S2a. The mutual inductance between the on-chip coil and nano-SQUIDs are extracted from the voltage-to-control-current curves scales down from $\sim$0.3 pH to $\sim$0.07 pH shown in of Fig.~S2b. The dashed line is calculated from an equation representing a mutual inductance between a line current 300 nm away from the edges of the nano-SQUIDs and the nano-SQUIDs as indicated inside of Fig.~S2b. The effective magnetic capture area scales down from $\sim$4~$\mu$m$^2$ to $\sim$~0.28 $\mu$m$^2$, as shown in Fig.~S2b, which follows a square law of the size for the SQUID loop.

\subsection{Smaller nano-SQUIDs}

Nano-SQUIDs with 50-nm and 10-nm square SQUID loop were also made.
Very large modulation voltage was observed as 0.8 mV at 4.2 K and remained as 0.2 mV at 40 K in the nano-SQUID with 50 nm square SQUID loop, as shown in Fig.~S3. The fluctuation of the curves may be caused by electronic or mechanical noise from physical property and measurement system (PPMS).

\subsection{Noise properties}
The optimal size of the square SQUID loop was found to be $\sim$400~nm, where the coupling of the on-chip coil is strong, the output voltage is high, the FHIB direct-write time is short (less than 1 minute), and the flux noise is small. The flux noise of a nano-SQUID was measured by an open loop method. The nano-SQUID with a 400-nm square SQUID loop and a junction of 200 nm at 100 kHz was measured to have a flux noise of ~300~n$\Phi_0$/Hz$^{-1/2}$ at 4.2 K, as shown in Fig.~S4a. The black curve represents the background noise of the measurement system including the pre-amplifier. The flux noise at 100 kHz was also plotted versus a varying flux bias together with the \textit{V-B} curve in Fig.~S4b.

\subsection{Characterizations of the tran-impedance amplifier}
Nano-SQUIDs with 400-nm square SQUID loop of $\sim$~50 $\Omega$ with better impedance matching was made to transfer the current signal in on-chip coil to output voltage. The \textit{I-V} characteristics from 4.2 K up to 40 K are shown in Fig.~S5a. The $I_\text{C}$ decreased from 12 $\mu$A to 1 $\mu$A, while the $R_\text{N}$ decreased from $\sim$56~$\Omega$ to $\sim$45~$\Omega$. The voltage-to-control-current curves at different temperatures from 4.2 K up to 30 K are present in Fig.~S5b. The modulation voltages are 0.5 mV, 0.3 mV, 0.17 mV, and 0.03 mV, respectively.

\begin{addendum}

 \item[Authors contribution] H.L. and S.C. conceived and planned the experiments. H.C designed and prepared the chips by using laser photolithography, argon ion milling and gold etching.  H.L, and E.Y.C. made the chips by using helium ion microscope. H.L., H.C., and E.Y.C carried out the measurement and characterization at low temperature. H.L, E.Y.C. and S.C. contributed to the interpretation of the results. S.C. took the lead in writing the manuscript.
 
 \item The authors thank Yan-Ting Wang and Stephen McCoy for the morphological analysis of the sample by using atomic force morphology and scanning electronic morphology, Shozo Yoshizumi of Quantum Design for assistance with PPMS measurements, and Miranda Vinay for help with proofreading the manuscript. This work was supported by the Air Force Office of Scientific Research under Grants FA955015-1-0218, and FA9550-19-C-0001; the National Science Foundation under Grant 1664446; the University of California Office of the President, Multi-campus Research Programs and Initiatives under Award No.009556-002; and by the Army Research Office Grant W911NF1710504. 
 
\item[Competing Interests] The authors declare that they have no
competing financial interests.
\item[Correspondence] Correspondence and requests for materials
should be addressed to Shane A. Cybart~(email: cybart@ucr.edu).
\end{addendum}

\begin{figure}
    \centering
    \includegraphics[width=3 in]{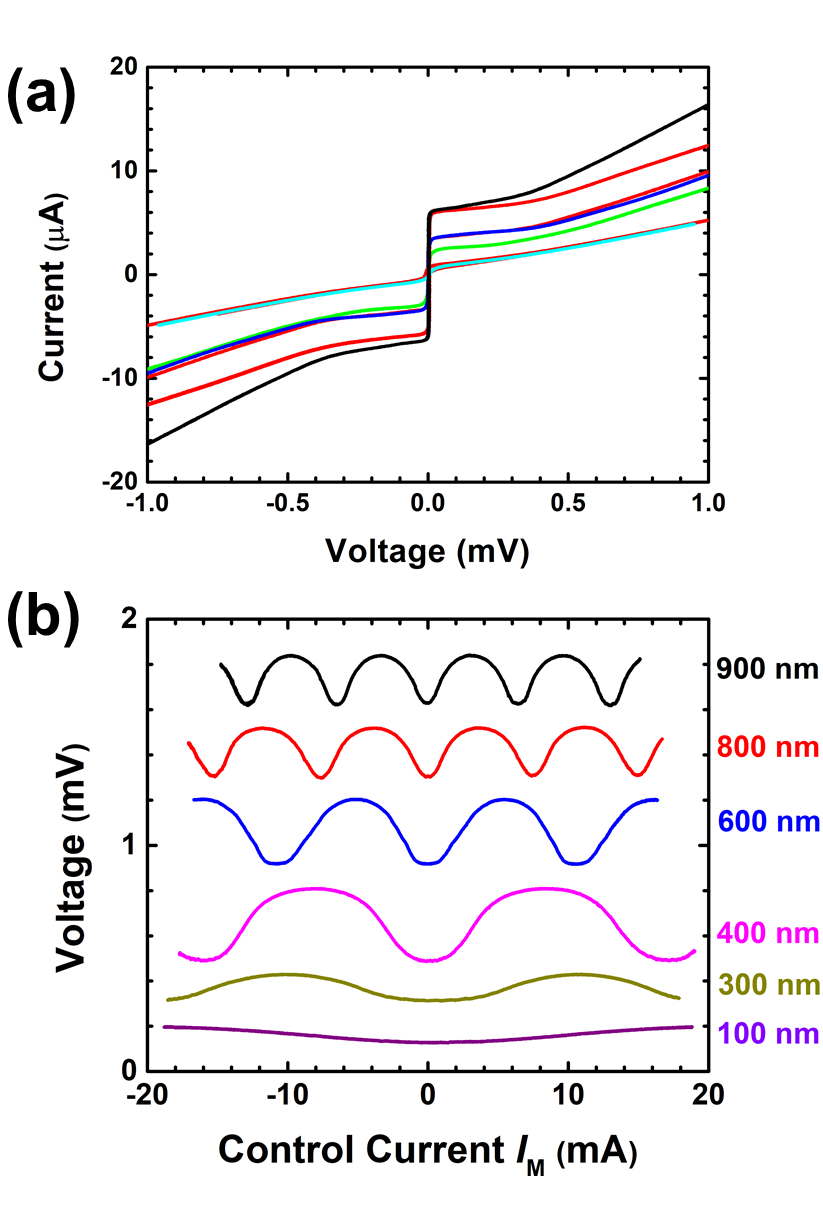}
    \caption{(a) \textit{I-V} characteristics of the nano-SQUIDs with varying size at 4.2~K. (b) The Voltages across the nano-SQUIDs oscillate with the current $I_\text{M}$ swept through the on-chip control line at 4.2~K. Some of the curvs on y-axis were offset to avoid overlap.}
       \label{fig:fig2a}
\end{figure}

\begin{figure}
    \centering
    \includegraphics[width=3 in]{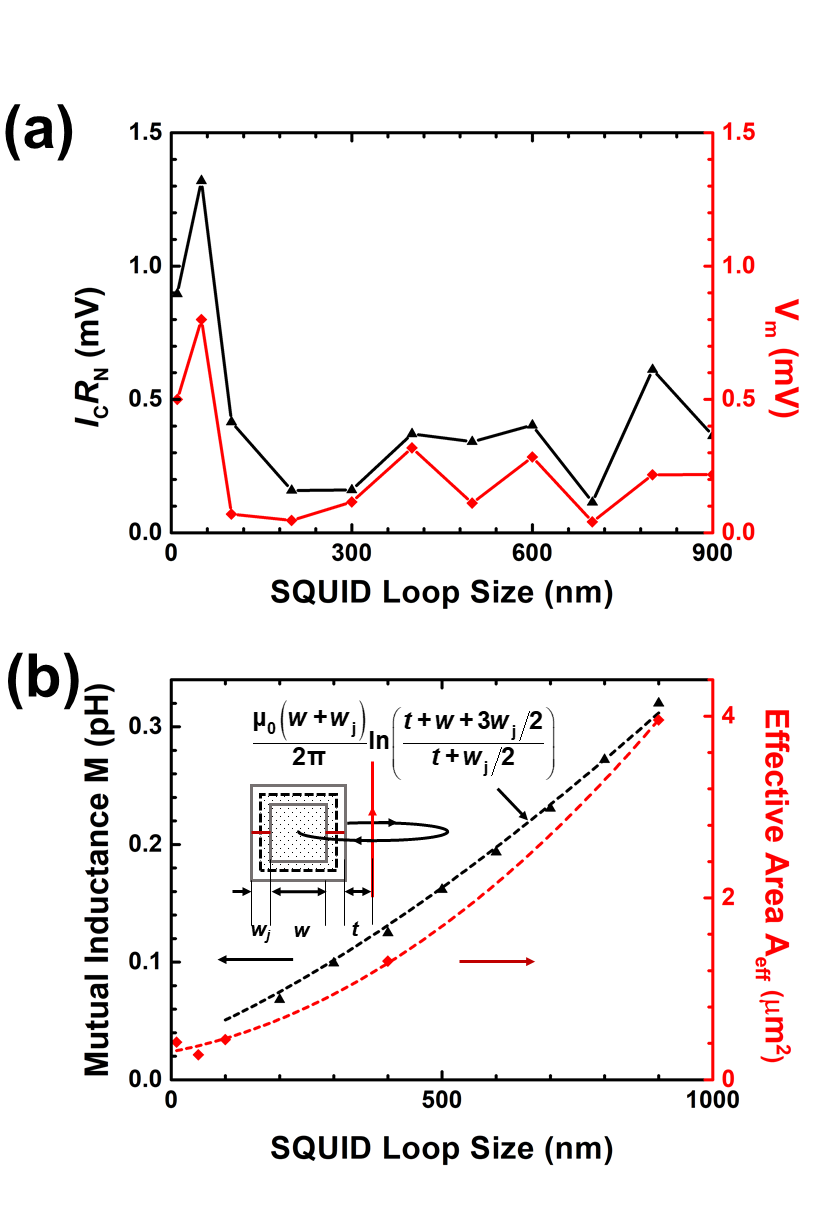}
    \caption{(a) The voltage modulation and $I_\text{C}R_\text{N}$ product of the nano-SQUIDs. (b) The mutual inductance between the on-chip coil and nano-SQUIDs, and the effective area decreases as the size of center square SQUID loop decreases. }
       \label{fig:fig2a}
\end{figure} 

\begin{figure}
    \centering
    \includegraphics[width=3 in]{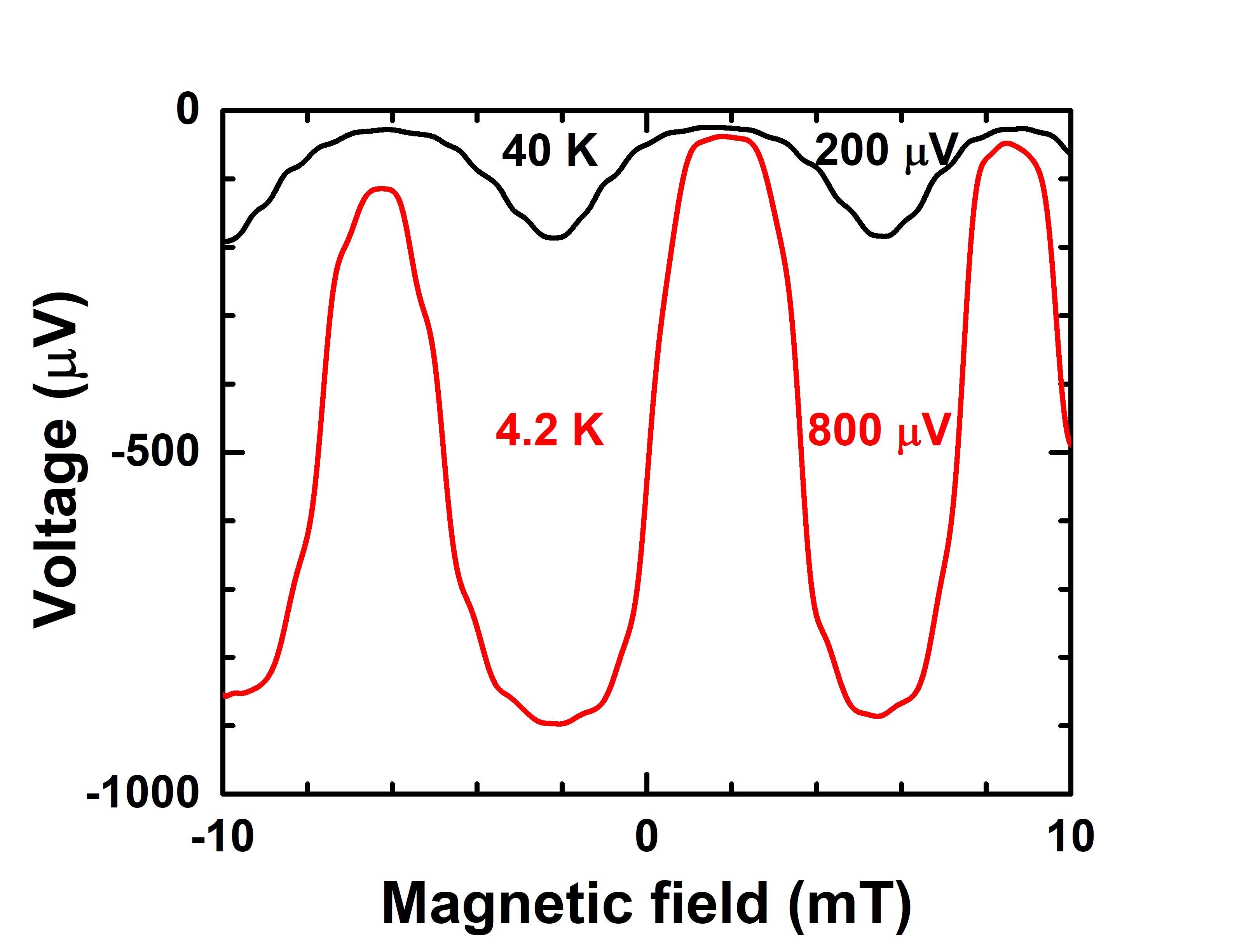}
       \caption{\textit{V-B} curves of the nano-SQUIDs with 50-nm wide square SQUID loop at 4.2 K and 40 K measured in a PPMS.}
       \label{fig:fig2a}
\end{figure}

\begin{figure}
    \centering
    \includegraphics[width=3 in]{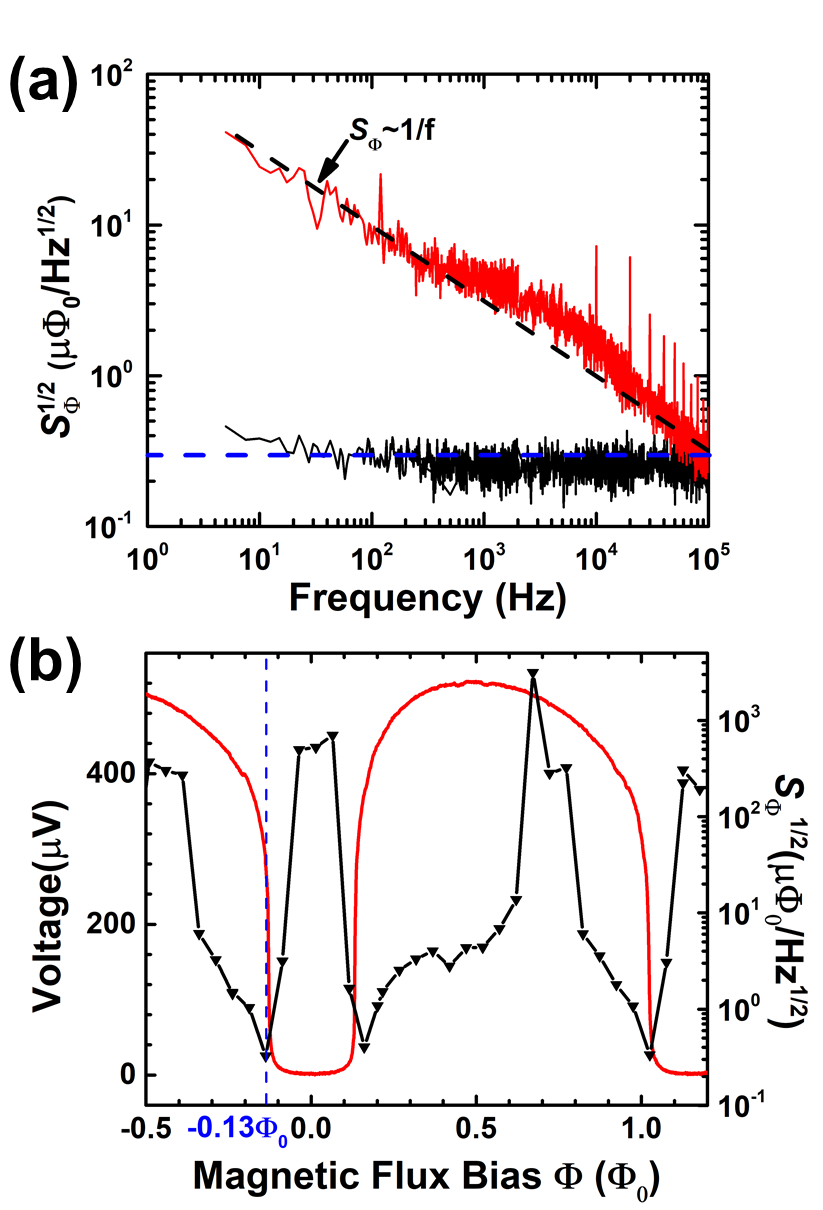}
    \caption{ (a) Flux noise spectrum $S_{\Phi_0}^{1/2}$ obtained from small signal methods for the nano-SQUID with 400-nm wide square SQUID loop measured at a bias current of 50.5 $\mu$A and optimal flux bias (magnetic flux bias of -0.13$\Phi_0$).The black curve is the background noise of the measurement system including the pre-amplifier.  The black dashed line is $S_{\Phi}$ $\sim$ 1/\textit{f}. (b)Voltage as a function of magnetic flux bias and the flux noise spectrum at 100 kHz. The optimal flux bias is indicated by the blue dashed line. }
       \label{fig:fig2a}
\end{figure}

\begin{figure}
    \centering
    \includegraphics[width=3 in]{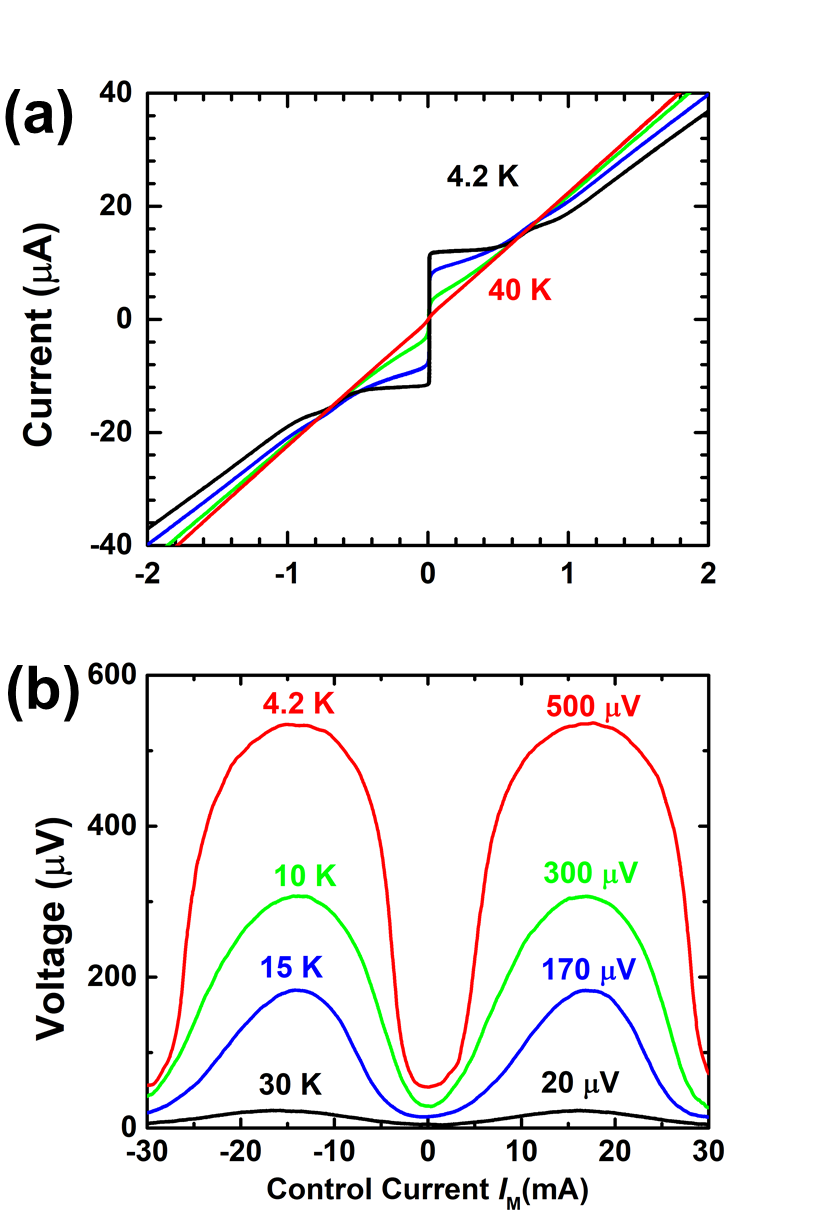}
    \caption{ (a) {I-V} characteristics of the nano-SQUIDs consisting of two 200-nm wide Josephson junctions and a 400-nm wide square SQUID loop at four different temperatures, 4.2 K, 15 K, 25 K and 40 K and. $I_C$ decreased from 12~$\mu$A to 1~$\mu$A, while the $R_N$ decreased from $\sim$56~$\Omega$ to $\sim$44~$\Omega$. (b) Voltage modulated by an on-chip control current at different temperatures, from 4.2 K, 10 K, 15 K, 30 K. The modulation voltage are 0.5 mV, 0.3 mV, 0.17 mV and 0.03 mV, respectively. }
       \label{fig:fig2a}
\end{figure}
